\documentclass[12pt,a4paper]{article}

\usepackage{amsmath,amssymb,graphicx,subcaption,xcolor}
\def\be{\begin{equation}}
\def\ee{\end{equation}}
\def\d{\text{d}}
\def\e{\text{e}}

\def\Sm{\Sigma_-}

\def\Sp{\Sigma_+}

\def\SN{\mathcal{N}}

\begin{document}
\begin{center}
{\Large{\bf Transition analysis of the non-OT $G_2$ stiff fluid spike solution}}

\

{\bf W C Lim$^1$ and M Z A Moughal$^{1,2}$}

\

${}^1$ Department of Mathematics, University of Waikato, Private Bag 3105, Hamilton 3240, New Zealand

\

${}^2$ Department of Basic Sciences and Humanities, College of Electrical and Mechanical Engineering, National University of Science and Technology, Islamabad, Pakistan

\

wclim@waikato.ac.nz, moughalzubair@gmail.com
\end{center}

\begin{abstract}
We use the technique developed in Moughal's doctoral thesis to analyse the joint spike transition,
revealing new groups of worldlines which undergo distinct transitions, and correcting misconceptions about spikes.
\end{abstract}


\section{Introduction}

The dynamics of a spacetime (for fluids with soft equation of state) as it evolves toward a generic spacelike singularity can be described as 
an infinite sequence of transitions between Kasner saddle states.
Neighbouring worldlines experience diverging sequences, giving the dynamics a chaotic nature.
Such transitions are described by vacuum Bianchi type II solutions.
This dynamics is commonly known as the Belinskii-Khalatnikov-Lifschitz (BKL) dynamics~\cite{art:LK63,art:BKL1970,art:BKL1982}.
A Kasner epoch refers to the time spent near a Kasner state.
Each Kasner state can be characterised by its BKL parameter $u$, which is computed from the expansion shear components.
The BKL parameter $u$ is a real number with $u \geq 1$, and it decreases by $1$ for each Bianchi type II transition, until
when $1<u<2$ before the transition, where $u$ is taken to $\frac{1}{u-1}$ at the next Kasner epoch.
A Kasner era refers to a sequence of Kasner epochs with decreasing BKL parameter $u$.

Furthermore, there are two possible ways to make a vacuum Bianchi type II transition, depending on the sign of a certain spatial curvature component.
If the spatial distribution of this variable has both signs, then a new dynamics occur in the neighbourhood of worldlines where this variable is zero.
Here, the sequence of vacuum Bianchi type II transition is replaced by a sequence of spike transitions. The new dynamics is called spike oscillations.
Spikes were first discovered numerically in 1993~\cite{art:BergerMoncrief1993} in the class of vacuum models called the general (unpolarised) Gowdy $T^3$ cosmologies, which admit two commuting spacelike Killing vector fields that act orthogonally transitively.
$T^3$ refers to the global spatial topology, which is a 3-torus, but its role in spike formation is not essential. We refer to this class of models (without specifying any global spatial topology) as orthogonally transitive (OT) $G_2$ models.
The orthogonal transitivity condition limits the sequence of transitions to a single Kasner era plus the first Kasner epoch of the next era, terminating at a final Kasner state or a final, permanent spike. Removing the orthogonal transitivity condition recovers the non-terminating BKL dynamics and uncovers a new spike transition in place of the permanent spike. This new spike transition (called joint spike transition~\cite{art:HeinzleUgglaLim2012}) is a more complex spike transition that begins in one Kasner era and ends in the next Kasner era. See~\cite{art:HeinzleUgglaLim2012} for a thorough introduction.

The exact orthogonally transitive (OT) $G_2$ (vacuum) spike solution was discovered in~\cite{art:Lim2008}.
In the paper~\cite{art:Limetal2009}, a matching procedure was developed and was used to provide numerical evidence that the exact solution describes a simple spike transition
that occurs within the same Kasner era.
It was not until the paper~\cite{art:Lim2015} that the exact non-OT $G_2$ (vacuum) spike solution (which could describe a joint spike transition) was discovered.
So far the solutions we mentioned are vacuum solutions. 
The Geroch transformation used to generate the non-OT $G_2$ spike solution can be applied to stiff fluid solutions,
which is of some interest.
The stiff fluid spike solution was given in the paper~\cite{art:ColeyLim2016}.
Although the method cannot be extended to other forms of source, it follows from the BKL conjecture that models with soft equation of state are asymptotic to the vacuum case.

While we can clearly describe the transition along the spike worldline and the transition along a typical (non-spike) worldline far away enough from the spike worldline, we often assume that the intermediate worldlines near the spike worldlines must be undergoing some sort of indistinct transition.
This turns out to be a misconception about spikes.
Recently in Moughal's doctoral thesis~\cite{thesis:Moughal2021,art:MoughalLim2021}, a new technique has been developed to analyse the transition times.
There, the technique was applied to cylindrical spikes, and it discovered new groups of intermediate worldlines that have distinct transition sequences.
In this paper, we shall use the technique to analyse the non-OT $G_2$ stiff-fluid spike solution, and discover similar groups of intermediate worldlines.

\section{The rotated Jacobs solution and the exact spike solution}

In this section we give the metric of the two exact solutions that we will be using -- the rotated Jacobs solution and the exact non-OT $G_2$ spike solution. The coordinate variables are $\tau$, $x$, $y$ and $z$, with $\tau \to \infty$ at the Big Bang singularity.
The rotated Jacobs solution is spatially homogeneous.
For the spike solution (with $K=0$), its spatial inhomogeneity is indicated by its dependence on the spatial coordinate $z$.
The reader can find the definitions of variables we use below in~\cite{art:HeinzleUgglaLim2012,art:Lim2015,art:ColeyLim2016,art:MoughalLim2021}.
Briefly, we shall represent the nonzero metric components in Iwasawa frame variables $N$, $b^1$, $b^2$, $b^3$, $n_1$, $n_2$ and $n_3$ as follows:
\begin{align}
	g_{00} &= - N^2,
\\
	g_{11} &= \e^{-2b^1},\quad g_{12} = \e^{-2b^1} n_1,\quad g_{13} = e^{-2b^1} n_3,
\\
	g_{22} &= \e^{-2b^2} + e^{-2b^1} n_1{}^2,\quad g_{23} = \e^{-2b^1} n_1 n_2 + \e^{-2b^2} n_3,
\\
	g_{33} &= \e^{-2b^3} + \e^{-2b^1} n_2{}^2 + \e^{-2b^3} n_3{}^2.
\end{align}

The Jacobs solution~\cite{art:Jacobs1968} is the stiff fluid Bianchi type I solution.
The rotated Jacobs solution (with $K=0$) is given by:~\cite[Equations (35)--(45)]{art:ColeyLim2016}
\begin{align}
\label{rotated_Jacobs}
	N &= -\e^{-\frac14(w^2+3+4\rho_0)\tau}
\\
	\e^{-2b^1} &= \lambda
\\
	\e^{-2b^2} &= \frac{\mathcal{A}^2}{\lambda}
\\
	\e^{-2b^3} &= \e^{-\frac12(w^2+3+4\rho_0)\tau} \mathcal{A}^{-2} 
\\
	n_1 &= \frac{n_{30} \e^{-(w-1)\tau} + n_{10} n_{20} \e^{-\frac12(w^2-1+4\rho_0)\tau}}{\lambda}
\\
  	n_2 &= \frac{n_{20} \e^{-\frac12(w^2-1+4\rho_0)\tau}}{\lambda}
\\
        n_3 &= \mathcal{A}^{-2} \left[ n_{10} \e^{-\frac12[(w-1)^2+4\rho_0]\tau} + n_{30} (n_{10} n_{30} - n_{20}) \e^{-\frac12[(w+1)^2+4\rho_0]\tau} \right],
\intertext{where}
\label{Asq}
        \mathcal{A}^2 &= \e^{-2\tau} + n_{10}^2 \e^{-\frac12[(w-1)^2+4\rho_0]\tau} + (n_{10} n_{30} - n_{20})^2 \e^{-\frac12[(w+1)^2+4\rho_0]\tau}
\\
\label{lambda}
        \lambda &= \e^{(w-1)\tau} + n_{20}^2 \e^{-\frac12(w^2-1+4\rho_0)\tau} + n_{30}^2 \e^{-(w+1)\tau}.
\end{align}
The stiff fluid density is
\be
	\rho = \rho_0 \e^{\frac12(w^2+3+4\rho_0)\tau}.
\ee
Setting $K=0$ implies
\be
\label{n20choice}
        n_{20} = \frac{4w}{(w+3)(w-1)+4\rho_0} n_{10} n_{30}.
\ee
The rotated Jacobs solution is simply the Jacobs solution in a rotating orthonormal frame. 
$w$ and $\rho_0$ specify the Jacobs solution, while $n_{10}$ and $n_{30}$ affect the timing of the two rotation transitions.
Setting $n_{10}$ and $n_{30}$ to zero returns the solution in a non-rotating frame.
If $\rho_0=0$, then the solution simplifies to the rotated Kasner solution~\cite[Equations (12)--(19)]{art:Lim2015}.

The exact non-OT $G_2$ stiff fluid spike solution was given by Equations (51)--(58) of the paper~\cite{art:ColeyLim2016}. In this paper we shall focus on the special case $K=0$.
We take this opportunity to correct a typographical error occurring in Equation (39) of~\cite{art:ColeyLim2016} and in Equation (16) of~\cite{art:Lim2015}.
\begin{align}
\label{nonOT_spike}
        N &= -\e^{-\frac14(w^2+3+4\rho_0)\tau} \sqrt{F}
\\
  	\e^{-2b^1} &= \lambda F^{-1}
\\
  	\e^{-2b^2} &= \mathcal{A}^2 \lambda^{-1} F
\\
\label{nonOT_spike_b3}
        \e^{-2b^3} &= \e^{-\frac12(w^2+3+4\rho_0)\tau} \mathcal{A}^{-2} F
\\
  	n_1 &= -2w(w-1) n_{30} z^2 + 2 (w-1) K y z - 2(w-1)\omega_0 z
\notag\\
        &\quad
              	+ \frac{\omega^2}{\lambda}(n_{30} \e^{-(w+1)\tau} + n_{10} n_{20} \e^{-\frac12(w^2-1+4\rho_0)\tau})
\notag\\
        &\quad  -\Bigg[ n_{30} w \e^{-2\tau} + n_{10} n_{20} \frac{(w+3)(w-1)+4\rho_0}{(w-1)^2+4\rho_0} \e^{-\frac12[(w-1)^2+4\rho_0]\tau}
\notag\\
        &\quad
              	+ n_{20} n_{30} (n_{10} n_{30} - n_{20}) \frac{(w-3)(w+1)+4\rho_0}{(w+1)^2+4\rho_0} \e^{-\frac12[(w+1)^2+4\rho_0]\tau} \Bigg]
\\
  	n_2 &= n_{20} \Bigg[ \frac{\omega^2}{\lambda} \e^{-\frac12(w^2-1+4\rho_0)\tau} - \frac{(w+3)(w-1)+4\rho_0}{(w-1)^2+4\rho_0} \e^{-\frac12[(w-1)^2+4\rho_0]\tau}
\notag\\
        &\quad
              	- n_{30}^2 \frac{(w-3)(w+1)+4\rho_0}{(w+1)^2+4\rho_0} \e^{-\frac12[(w+1)^2+4\rho_0]\tau} \Bigg]
\\
  	n_3 &= \mathcal{A}^{-2} \left[ n_{10} \e^{-\frac12[(w-1)^2+4\rho_0]\tau} + n_{30} (n_{10} n_{30} - n_{20}) \e^{-\frac12[(w+1)^2+4\rho_0]\tau} \right],
\label{nonOT_spike_n3}
\intertext{where}
\label{F}
	F &= \lambda^2 + \omega^2
\\
\label{omega}
        \omega &= 2w n_{30} z.
\end{align}
The stiff fluid density for the spike solution is
\be
	\rho = \rho_0 \e^{\frac12(w^2+3+4\rho_0)\tau} F^{-1}.
\ee
$w$ and $\rho_0$ specify the initial Jacobs equilibrium point in the state space, while $n_{10}$ and $n_{30}$ affect the timing of the transitions.

In both solutions, $\tau$ increases to infinity at the Big Bang singularity.

\section{Analysis of transition times}

To explain how the analysis works, we warm up with the rotated Jacobs solution first. 

\subsection{Double frame transitions in the rotated Jacobs solution}

The original, non-rotated Jacobs solution is a spatially homogeneous solution of the Einstein field equations.
The dynamics of such class of solutions can be described by a set of autonomous ordinary differential equations when formulated
in the orthonormal frame approach using expansion-normalised variables (see~\cite{book:WainwrightEllis1997} for background on orthonormal frame approach and expansion-normalisation).
The evolution of spatially homogeneous solutions can be represented as an orbit in the state space of the expansion-normalised variables.
Spatially homogeneous solutions that are also self-similar (solutions with a proper homothetic vector field) can be represented as equilibrium points in the state space~\cite[Sections 1.2.3 and 5.3.3]{book:WainwrightEllis1997}.
The non-rotated Jacobs solution is such an example.
The equilibrium points are important in that the state space orbits of solutions typically spend a long time near equilibrium points, and only spend a short time transitioning between equilibrium points.
Transitions are represented by orbits connecting equilibrium points.
The non-rotated Jacobs solution is actually a two-parameter family of solutions (parameterised by $w$ and $\rho_0$), 
represented by a unit disk of equilibrium points in the $(\Sp,\Sm)$ plane in the state space.
Among the expansion-normalised variables, only the Hubble-normalised expansion shear components $\Sp$ and $\Sm$ appear in this paper. They will be introduced below.

The rotated Jacobs solution introduces two degrees of frame rotations. The two rotations may occur in different orders or simultaneously.
This is called double frame transitions~\cite{art:HeinzleUgglaLim2012,art:HeinzleUgglaRohr2009,art:Hewittetal2003}.
See Figure~\ref{double_frame_transition_plot} below for the orbits representing the transitions.

Near an equilibrium point, the state space variables are almost constant.
The state space variables must transition from one constant to another during a transition between equilibrium points.
A simple example is the hyperbolic tangent function, which is a fractional function of two different time-dependent terms:
\be
	\text{tanh}(\tau) = \frac{\e^\tau - \e^{-\tau}}{\e^\tau+\e^{-\tau}}.
\ee
The hyperbolic tangent function spend most time nearly constant.
For $\tau$ approximately less than $-2$, say, the term $\e^{-\tau}$ is dominant while $\e^\tau$ is sub-dominant.
Transition happens around $\tau=0$ when the dominant term $\e^{-\tau}$ is overtaken by a previously sub-dominant term $\e^\tau$.
This example is useful to keep in mind when reading the rest of this paper.

The rotated Jacobs solution is more complicated.
Two convenient variables to investigate are Hubble-normalised shear variables $\Sp$ and $\Sp+\sqrt{3}\Sm$, 
which isolate the effect of $\mathcal{A}^2$ and $\lambda$ respectively.
The Hubble expansion scalar $H$ is given by (see~\cite[Appendix B]{art:HeinzleUgglaLim2012})
\be
	H = -\frac13 N^{-1} \partial_\tau (b^1+b^2+b^3),
\ee
and the Hubble-normalised lapse $\SN$ is given by
\be
	\SN = NH.
\ee
The Hubble-normalised shear components are given by
\begin{align}
	\Sigma_{33} &= -2\Sp =  - 1 - \SN^{-1} \partial_\tau b^3
\\
	\Sigma_{11} &= \Sp+\sqrt{3}\Sm = - 1 - \SN^{-1} \partial_\tau b^1
\end{align}
For the rotated Jacobs solution these simplify to
\begin{align}
	\Sp &= -1 + \frac14 \SN^{-1} \partial_\tau \ln \mathcal{A}^2
\\
	\Sp+\sqrt{3}\Sm &= -1 + \frac12 \SN^{-1} \partial_\tau \ln \lambda,
\end{align}
where Hubble-normalised $\SN$ is constant:
\be
	\SN = NH = -\frac{1}{12}(w^2+3+4\rho_0).
\ee
We can see that $\partial_\tau \ln \mathcal{A}^2$ determines the transition times of $\Sp$, while $\partial_\tau \ln \lambda$ determines the transition times of $\Sp+\sqrt{3}\Sm$.
Both $\mathcal{A}^2$ and $\lambda$ have three terms, so each has three equilibrium states and two transitions.

\begin{figure}[t]
	\begin{center}
                \includegraphics[width=12cm]{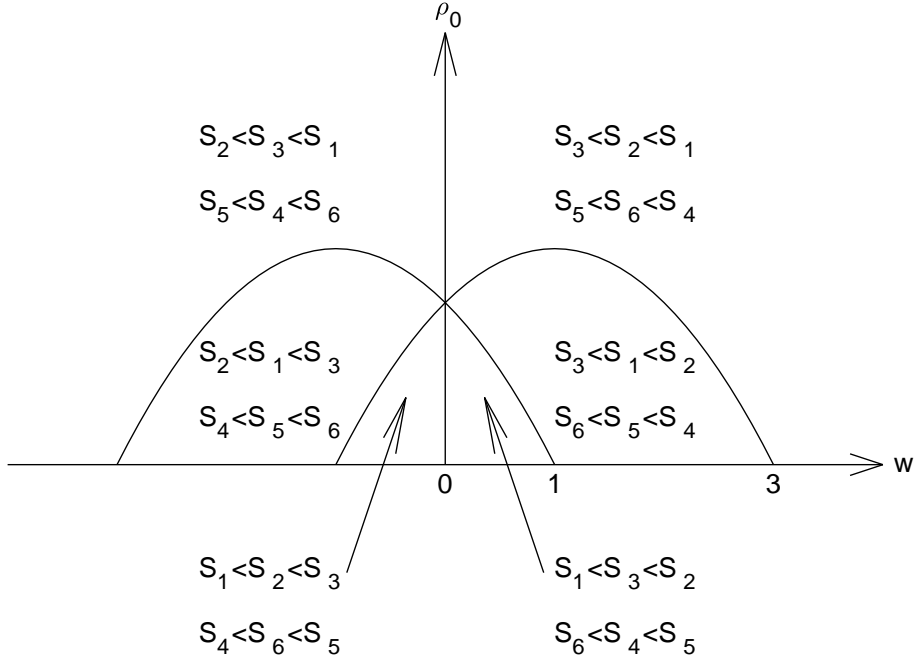}
		\caption{Order of dominance among $\{S_1,S_2,S_3\}$ and among $\{S_4,S_5,S_6\}$, depending on the parameters $w$ and $\rho_0$. The parabolas are given by~(\ref{parabolas}).}
		\label{fig:order_of_dominance}
	\end{center}
\end{figure}

For $\mathcal{A}^2$ in~(\ref{Asq}), we label its terms as follows:
\be
	T_1 = \e^{-2\tau},\quad T_2 = n_{10}^2 \e^{-\frac12[(w-1)^2+4\rho_0]\tau},\quad T_3 = (n_{10} n_{30} - n_{20})^2 \e^{-\frac12[(w+1)^2+4\rho_0]\tau},
\ee
and call the coefficients in their exponents
\be
	S_1 = -2,\quad S_2 = -\frac12[(w-1)^2+4\rho_0],\quad S_3 = -\frac12[(w+1)^2+4\rho_0].
\ee
Similarly, for $\lambda$ in~(\ref{lambda}), we label its terms
\be
	T_4 = \e^{(w-1)\tau},\quad T_5 = n_{20}^2 \e^{-\frac12(w^2-1+4\rho_0)\tau},\quad T_6 = n_{30}^2 \e^{-(w+1)\tau},
\ee
and the coefficients in their exponents
\be
	S_4 = w-1,\quad	S_5 = -\frac12(w^2-1+4\rho_0),\quad S_6 = -(w+1).
\ee
By comparing the exponent coefficients, we obtain the inequalities
\begin{align}
	S_1 > S_2 \ \text{and} \ S_6 > S_5 \ &\text{iff} \ \rho_0 > 1 - \tfrac14(w-1)^2,
\\
	S_1 > S_3 \ \text{and} \ S_4 > S_5 \ &\text{iff} \ \rho_0 > 1 - \tfrac14(w+1)^2,
\\
	S_2 > S_3 \ \text{and} \ S_4 > S_6 \ &\text{iff} \ w > 0.
\end{align}
We summarise this in Figure~\ref{fig:order_of_dominance}, which shows the order of dominance of the exponents. The two parabolas in the figure are given by
\be
\label{parabolas}
	\rho_0 = 1 -\tfrac14(w\pm1)^2.
\ee
Both the rotated Jacobs solution and the non-OT spike solution are multiply-represented, in the sense that the same state-space orbit yields multiple values of $w$~\cite{art:Lim2015}.
We shall present the transition times analysis using the case
\be
\label{parameter_case}
	w > 0,\quad \rho_0 > 1- \tfrac14(w-1)^2,
\ee
as Figure~\ref{fig:order_of_dominance} shows that it covers all possible values of $\rho_0$. In this case, we have
\be    
	S_3 < S_2 < S_1,\quad S_5 < S_6 < S_4.
\ee
That is, for $\mathcal{A}^2$, $T_3$ dominates at early $\tau$ (away from Big Bang singularity), while $T_1$ dominates at late $\tau$ (closer to Big Bang singularity). 
$T_2$ may dominate at intermediate time if its coefficient $n_{10}^2$ is large enough, with the exact condition to be determined by the analysis below.
Similarly, for $\lambda$, $T_5$ dominates early, while $T_4$ dominates late, with $T_6$ possibly dominating at intermediate time if its coefficient $n_{30}^2$ is large enough.
The analysis can be done similarly for the other cases.

The graph of 
\be
	\partial_\tau \ln \mathcal{A}^2 = \frac{S_1 T_1 + S_2 T_2 + S_3 T_3}{T_1 + T_2 + T_3}
\ee
has a cascading shape (By this we mean the function is monotone in $\tau$, with short transitions between long nearly-constant periods. Its graph resembles a series of waterfalls. See plots for $\Sp$ and $\Sigma_{11}$ in Figures~\ref{double_frame_transition_plot} and~\ref{double_frame_transition_plot_second_example}, and the plots in Figures~\ref{waterfall_cell1} and~\ref{waterfall_cell2} below.), with
\be
	\partial_\tau \ln \mathcal{A}^2 \approx
	\begin{cases}
	S_3=-\frac12[(w+1)^2+4\rho_0]	&\text{when $T_3$ dominates}\\
	S_2=-\frac12[(w-1)^2+4\rho_0]	&\text{when $T_2$ dominates}\\
	S_1= -2				&\text{when $T_1$ dominates.}
	\end{cases}
\ee
Similarly, the graph of
\be
	\partial_\tau \ln \lambda = \frac{S_4 T_4 + S_5 T_5 + S_6 T_6}{T_4 + T_5 + T_6}
\ee
also has a cascading shape, with
\be
        \partial_\tau \ln \lambda \approx
        \begin{cases}
        S_5=-\frac12(w^2-1+4\rho_0)		&\text{when $T_5$ dominates}\\
        S_6=-(w+1)       			&\text{when $T_6$ dominates}\\
        S_4=(w-1)       			&\text{when $T_4$ dominates.}
        \end{cases}
\ee
One can clearly see how these are slightly more complicated than the hyperbolic tangent function.
The transition time between two states is defined as the time when two terms are equal in magnitude.
Let the indices of the transition time $\tau_{AB}$ denote the states involved. For example, the transition time $\tau_{32}$ is obtained by solving $|T_3|=|T_2|$ for $\tau$.
The transition times for $\mathcal{A}^2$ are:
\begin{align}
        \tau_{32} &= \frac{\ln|n_{10}n_{30}-n_{20}|-\ln|n_{10}|}{w}
\\
        \tau_{21} &= \frac{4\ln|n_{10}|}{(w-3)(w+1) + 4\rho_0}
\\
\label{tau31}
        \tau_{31} &= \frac{4\ln|n_{10}n_{30}-n_{20}|}{(w+3)(w-1) + 4\rho_0}.
\end{align}
We call a sequence of equilibrium states (represented by the dominant term) a scenario.
If $\tau_{32} < \tau_{21}$, then we have the scenario $T_3 \rightarrow T_2 \rightarrow T_1$ (as $\tau$ increases), otherwise we have the scenario $T_3 \rightarrow T_1$.
$\tau_{32} < \tau_{21}$ implies
\be
\label{Asq_3states}
         \ln|n_{30}| < \frac{4 w}{(w-3)(w+1) + 4\rho_0}\ln |n_{10}| + \ln \left[ \frac{(w+3)(w-1) + 4\rho_0}{(w-3)(w+1) + 4\rho_0} \right].
\ee
The transition times for $\lambda$ are:
\begin{align}
\label{tau56}
	\tau_{56} &= \frac{4 (\ln |n_{20}|-\ln|n_{30}|)}{(w-3)(w+1) + 4\rho_0}
\\
\label{tau64}
	\tau_{64} &= \frac{\ln|n_{30}|}{w}
\\
	\tau_{54} &= \frac{4 \ln |n_{20}|}{(w+3)(w-1) + 4\rho_0}.
\end{align}
If $\tau_{56} < \tau_{64}$, then we have the scenario $T_5 \rightarrow T_6 \rightarrow T_4$, otherwise we have the scenario $T_5 \rightarrow T_4$.
$\tau_{56} < \tau_{64}$ implies
\be
\label{lambda_3states}
        \ln|n_{30}| > \frac{4 w}{(w-3)(w+1) + 4\rho_0} \left[ \ln \frac{4 w}{(w+3)(w-1) + 4\rho_0} + \ln |n_{10}| \right].
\ee
The right hand side of (\ref{lambda_3states}) is smaller than the that of (\ref{Asq_3states}), so it is possible to satisfy both (\ref{Asq_3states}) and (\ref{lambda_3states}).

Graphically, we want to produce orbits that show visually distinct double frame transitions. 
To do so we need one more concept -- transition duration. 
Transition duration is the time it takes to complete a transition, and is inversely proportional to the difference between the exponent coefficients of the two relevant terms.
The bigger the difference, the shorter the duration.
The exponent coefficients $S_i$ are determined by parameters $w$ and $\rho_0$. A good choice is $w=5$, $\rho_0=0$.
Bigger $w$ means bigger difference and shorter duration, giving visually more distinct transitions.

Taking into account the transition duration, one needs to choose $\ln|n_{30}|$ much bigger or smaller than the critical value to see distinct three states, which is not possible,
so we never see distinct three states in both $\lambda$ and $\mathcal{A}^2$.

\begin{figure}[t]
        \begin{center}
        \includegraphics[width=12cm]{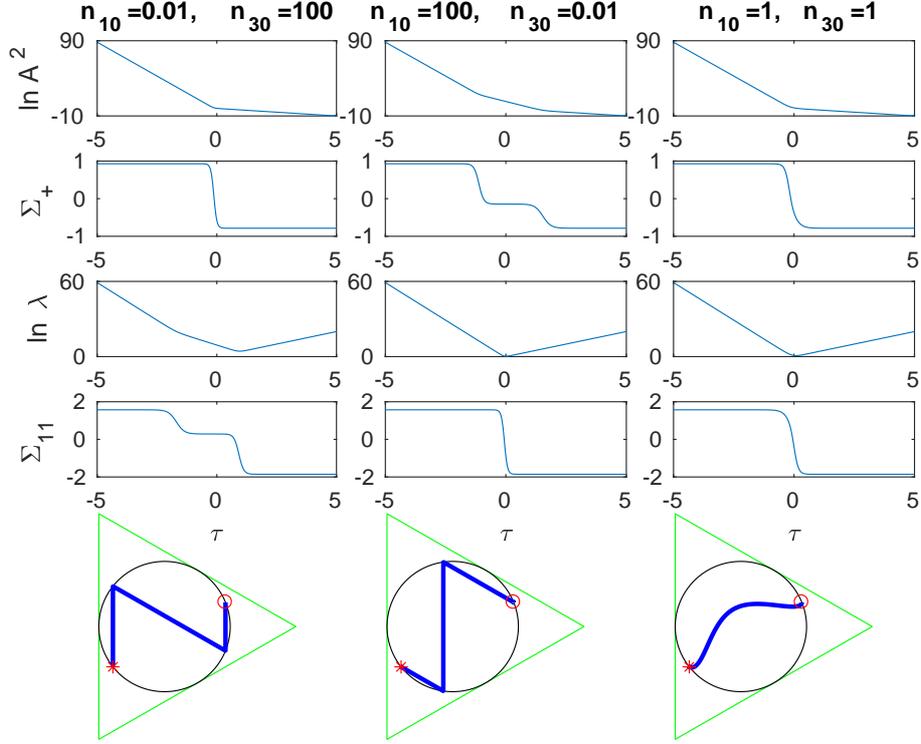}
        \caption{Plots of $\ln \mathcal{A}^2$, $\Sp$, $\ln \lambda$, $\Sigma_{11}$ and state space orbits projected onto the $(\Sp,\Sm)$ plane for three examples with $w=5$, $\rho_0=0$.
		The first two cases give visually distinctive double frame transitions. The third case gives visually indistinct ones.
		A red circle marks the start of the orbit, a red star marks the end.}
        \label{double_frame_transition_plot}
        \end{center}
\end{figure}

Next, we look at how the parameters $n_{10}$, $n_{20}$, $n_{30}$ affect the transition times.
It is clear from the transition times that larger numerator means later transition times.
From (\ref{lambda_3states}), one needs to choose $|n_{30}| \gg |n_{10}|$ to have visually distinct three states in $\lambda$ (and two states in $\mathcal{A}^2$). 
Similarly, from (\ref{Asq_3states}), one needs to choose $|n_{30}| \ll |n_{10}|$ to have visually distinct three states in $\mathcal{A}^2$ (and two states in $\lambda$). 
Either choices will give visually distinctive double frame transitions.
Choosing $|n_{30}| \approx |n_{10}|$ leads to two states visually in both $\lambda$ and $\mathcal{A}^2$, and yield visually indistinct double frame transitions.
Figure~\ref{double_frame_transition_plot} show an example for each case, with $w=5$, $\rho_0=0$, and
$n_{10}=0.01$, $n_{30}=100$ for the first case, 
$n_{10}=100$, $n_{30}=0.01$ for the second,
and
$n_{10}=1$, $n_{30}=1$ for the third.
The first case shows two $R_3 = -\Sigma_{12}$ frame transitions sandwiching an $R_1 = -\Sigma_{23}$ frame transition;
the second case shows two $R_1$ transitions sandwiching an $R_3$ transition;
the third case shows both $R_1$ and $R_3$ transitions occuring almost simultaneously.
Compare with Figure 4 of~\cite{art:Hewittetal2003} and Figure 4a of~\cite{art:HeinzleUgglaRohr2009}.
A second set of examples with $w=0.5$, $\rho_0=2$ is given in Figure~\ref{double_frame_transition_plot_second_example}.

\begin{figure}[t]
        \begin{center}
        \includegraphics[width=12cm]{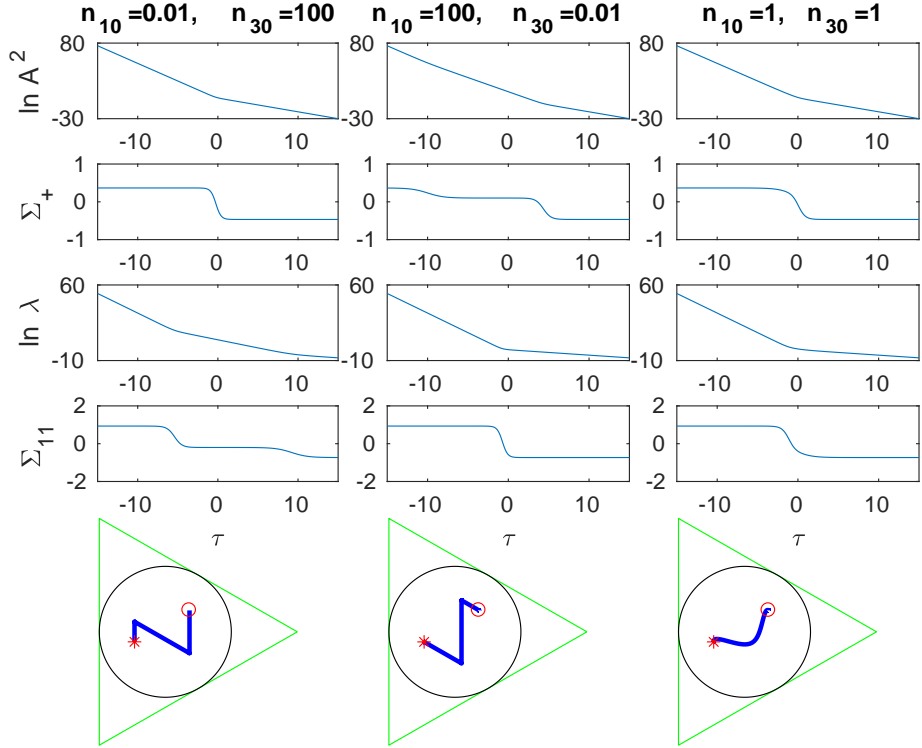}
        \caption{Plots of $\ln \mathcal{A}^2$, $\Sp$, $\ln \lambda$, $\Sigma_{11}$ and state space orbits projected onto the $(\Sp,\Sm)$ plane for three examples with $w=0.5$, $\rho_0=2$.
                The first two cases give visually distinctive double frame transitions. The third case gives visually indistinct ones.
                A red circle marks the start of the orbit, a red star marks the end.}
        \label{double_frame_transition_plot_second_example}
        \end{center}
\end{figure}

In this warm-up, we see the benefit of isolating the functions responsible for the transitions. In this case, they are $\lambda$ and $\mathcal{A}^2$.
They are a sum of different time-dependent terms, and derivative of their logarithm has a cascading graph reflecting the transitions between equilibrium states.
We are able to determine how to produce visually distinctive double frame transitions by careful choice of the parameter values.

\subsection{The joint spike transition in the spike solution}

Now turn to the spike solution. We focus on
\begin{align}
        \Sp &= -1 + \frac14 \SN^{-1} \partial_\tau \ln \mathcal{A}^2
\\
        s = \frac{\Sm+\sqrt{3}}{\Sp+1} &= \frac{ 2 \partial_\tau \ln \lambda - \partial_\tau \ln \mathcal{A}^2 - (w^2+3+4\rho_0)}{\sqrt{3} \partial_\tau \ln \mathcal{A}^2 }, 
\end{align}
where
\be
        \SN = NH = \frac16 \partial_\tau \ln F -\frac{1}{12}(w^2+3+4\rho_0).
\ee
The variable $s$ is convenient, in that it is spatially independent.
Only $F$ has spatial dependence, through the $z$-dependent $\omega$ (see (\ref{F}) and (\ref{omega})).
$s$ depends on both $\lambda$ and $\mathcal{A}^2$, whose dynamics has been analysed in the previous subsection.
We conclude that $s$ has up to 4 visually distinct states, and there are two different sets of these.
One set corresponds to the joint spike transition, which is our main focus. 
The other set is essentially a simple spike transition sandwiched by two $R_1$ frame transitions~\cite{art:Lim2015}.

We now analyse $\partial_\tau \ln F$. Introduce the time-independent (but spatial coordinate $z$-dependent) term 
\be
	T_7 = \omega = 2wn_{30}z,
\ee
with trivial exponential coefficient $S_7=0$, so that
\be
	F= \lambda^2 + \omega^2 = (T_4 + T_5 + T_6)^2 + T_7^2.
\ee
The graph of 
\be
	\partial_\tau \ln F = \frac{2(T_4 + T_5 + T_6)(S_4 T_4 + S_5 T_5 + S_6 T_6) + 2 S_7 T_7}{(T_4 + T_5 + T_6)^2 + T_7^2}
\ee
has a cascading shape, with
\be
        \partial_\tau \ln F \approx
        \begin{cases}
        2S_5=-(w^2-1+4\rho_0)	&\text{when $T_5$ dominates}\\
        2S_6=-2(w+1)		&\text{when $T_6$ dominates}\\
	2S_7=0			&\text{when $T_7$ dominates}\\
        2S_4=2(w-1)		&\text{when $T_4$ dominates.}
        \end{cases}
\ee
We have to consider two cases: the case $w > 1$ (in which $S_4>S_7$) and the case $w < 1$ (in which $S_4<S_7$).

\begin{figure}[t]
	\begin{center}
        \includegraphics[width=12cm]{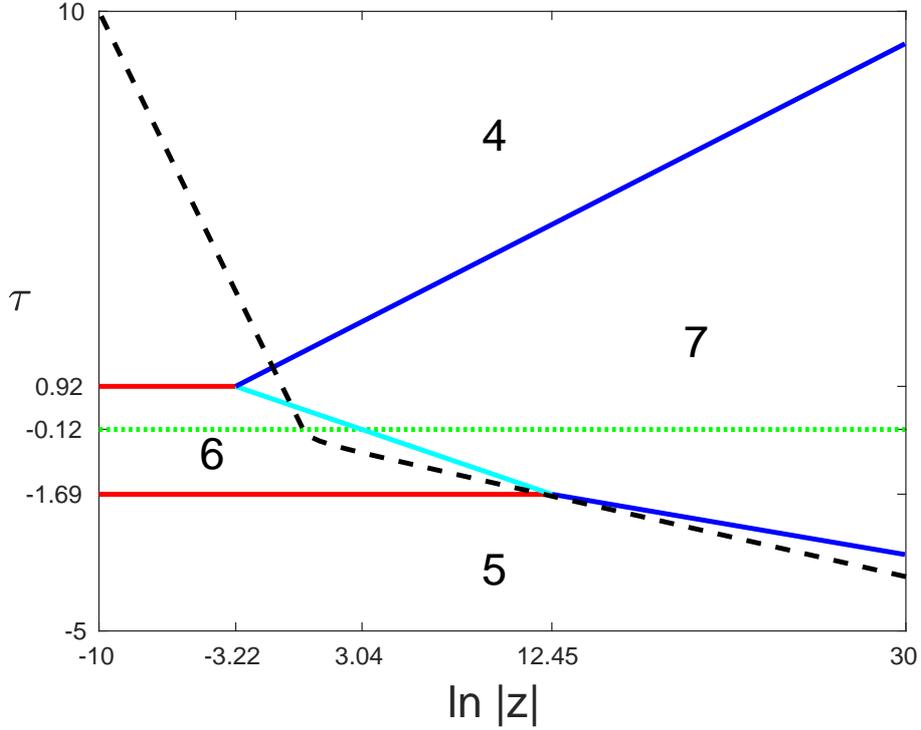}
        \caption{Plot of the cells and their boundaries -- transition times (solid lines) for $F$ for Case $w>1$, showing the different scenarios along each fixed $|z|$.
		Each cell is labelled with the index of the dominant term. The parameter values used here are $w=5$, $\rho_0=0$, $n_{10} = 0.01$, $n_{30} = 100$.
		$\ln |z|_1 \approx -3.22$, $\ln |z|_2 \approx 3.04$, $\ln |z|_3 \approx 12.45$, $\tau_{56} \approx -1.69$, $\tau_{31} \approx -0.12$, $\tau_{64} \approx 0.92$.
		Dashed line indicates the particle horizon. Recall that the transition time in $\mathcal{A}^2$ is $\tau_{31}$, and the transition times in $\lambda$ are $\tau_{56}$ and $\tau_{64}$.}
        \label{cell1}
	\end{center}
\end{figure}

\begin{figure}[t]
        \begin{center}
        \includegraphics[30,150][400,250]{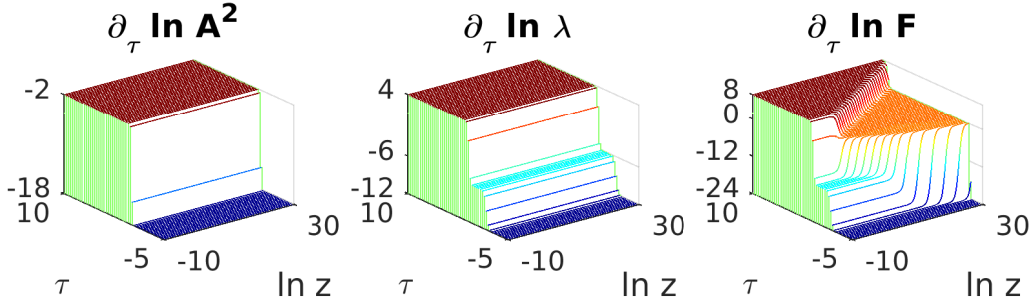}
        \caption{The cascading graphs of $\partial_\tau \ln \mathcal{A}^2$, $\partial_\tau \ln \lambda$ and $\partial_\tau \ln F$, with the same parameter values used in Figure~\ref{cell1}.}
        \label{waterfall_cell1}
        \end{center}
\end{figure}

\subsubsection{Case $w>1$}

For the case $w > 1$, we have $S_5 < S_6 < S_7 < S_4$.
Focussing on the joint spike transition, along worldlines with large enough $|z|$, $T_7$ will become dominant before $T_6$ does, so the scenario in $F$ is $T_5 \to T_7 \to T_4$.
Worldlines with large $|z|$ make the first transition, which happens when $|T_5| = |T_7|$, giving the transition time
\be
        \tau_{57} = -\frac{2}{w^2-1+4\rho_0} \ln \left( \frac{|\omega|}{n_{20}^2}\right).
\ee
Larger $|z|$, smaller $|n_{10}|$ or $|n_{30}|$ all give earlier transition time (more negative $\tau_{57}$).
Worldlines with large $|z|$ also make the last transition, which happens when $|T_7| = |T_4|$, giving the transition time
\be
        \tau_{74} = \frac{\ln|\omega|}{w-1}.
\ee
Larger $|z|$ or $|n_{30}|$ give later transition time (larger $\tau_{74}$).

In the other extreme, along worldlines with small enough $|z|$, $T_7$ is always sub-dominant, so $\partial_\tau \ln F$ behaves like $\partial_\tau \ln \lambda$,
with the scenario $T_5 \to T_6 \to T_4$.
For intermediate $|z|$, the scenario in $F$ is $T_5 \to T_6 \to T_7 \to T_4$,
with 
\be
	\tau_{67} = - \frac{\ln|\frac{2wz}{n_{30}}|}{w+1}.
\ee
The boundary between small and intermediate $|z|$ is found by solving $\tau_{64}=\tau_{74}$ (or $\tau_{64}=\tau_{67}$), giving
\be
	|z|_1=\frac{|n_{30}|^{-\frac{1}{w}}}{2w}.
\ee
The boundary between large and intermediate $|z|$ is found by solving $\tau_{56}=\tau_{57}$ (or $\tau_{56}=\tau_{67}$), giving
\be
        |z|_3=\frac{|n_{30}|}{2w} \left|\frac{n_{30}}{n_{20}}\right|^{\frac{4(w+1)}{(w-3)(w+1)+4\rho_0}}.
\ee
This completes the scenarios in $F$. We summarise them in Figure \ref{cell1}.
The parameter values used here are
\be
 	w=5,\ \rho_0=0,\ n_{10} = 0.01,\ n_{30} = 100, 
\ee
giving
\be
	 \ln |z|_1 \approx -3.22,\ \ln |z|_3 \approx 12.45,\ \tau_{56} \approx -1.69,\ \tau_{64} \approx 0.92.
\ee
The transition times partition the spacetime into cells. Each cell is represented by an equilibrium point (in this case, a Kasner equilibrium point).
Dashed line indicates the particle horizon (see~(\ref{particle_horizon}) in Appendix).
This diagram shows that the width of the spike (characterised by the $z$-dependent transition times) is super-horizon most of the time, and is only sub-horizon around $\tau_{64} \approx 0.92$.
Also plotted in the figure is the $R_1$ frame transition time occurring at $\tau_{31} \approx -0.12$, which is roughly the time when the particle horizon bends.
The cascading graphs of $\partial_\tau \ln \mathcal{A}^2$, $\partial_\tau \ln \lambda$ and $\partial_\tau \ln F$ are plotted in Figure~\ref{waterfall_cell1}.

\begin{figure}[t]
        \begin{center}
        \includegraphics[height=8cm]{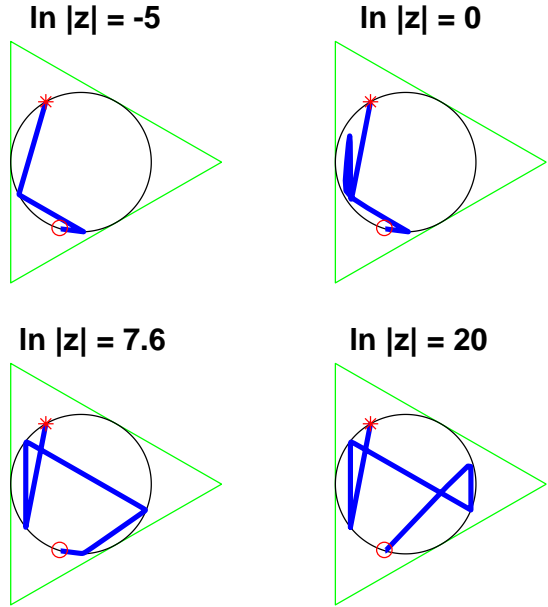}
        \caption{The state space orbits projected onto the Hubble-normalised $(\Sp,\Sm)$ plane, showing a distinctive orbit for each of the four groups of worldlines.
                The parameter values used here are $w=5$, $\rho_0=0$, $n_{10} = 0.01$, $n_{30} = 100$. The representative worldlines used are
		$\ln |z| = -5,\ 0,\ 7.6,\ 20$. A red circle marks the start of the orbit, a red star marks the end.
                }
        \label{orbits}
        \end{center}
\end{figure}

The above analysis reveals more details about spike transitions and clarifies previously held misconceptions about transient spikes.
Previously, we only made statements about the transitions along two groups of worldlines -- the spike worldline $z=0$ and along faraway worldlines (without saying how far is faraway), 
while not much is said about the intermediate worldlines near the spike worldline.
We also incorrectly thought that the spike worldline is the sole member of its group.
The diagram shows that there are four typical groups of worldlines that can have distinct transitions.
Worldlines with $|z| < |z|_1$ undergo the same sequence of transitions as the spike worldline, namely
\be
	\tau_{56},\ \tau_{31},\ \tau_{64}.
\ee
The analysis also quantifies what counts as faraway -- namely $|z| > |z|_3$. These worldlines undergo
\be
	\tau_{57},\ \tau_{56},\ \tau_{31},\ \tau_{64},\ \tau_{74}.
\ee
$\tau_{67}$ occurs earlier than $\tau_{31}$ for
\be
	|z| > |z|_2 = \frac{ |n_{30}|}{2w} \left| \frac{n_{10}}{n_{10}n_{30}-n_{20}}\right|^{\frac{w+1}{w}}.
\ee
This gives rise to two different scenarios for the intermediate worldlines with $|z|_1 < |z| < |z|_3$ -- those with $|z|_1 < |z| < |z|_2$ undergo the sequence of transitions in the following order
\be
	\tau_{56},\ \tau_{31},\ \tau_{67},\ \tau_{64},\ \tau_{74},
\ee
while those with $|z|_2 < |z| < |z|_3$ undergo
\be
        \tau_{56},\ \tau_{67},\ \tau_{31},\ \tau_{64},\ \tau_{74}.
\ee
In Figure~\ref{cell1}, $\ln |z|_2 \approx 3.04$.

Figure~\ref{orbits} show the distintive orbits for each of the four groups of worldlines.
Bear in mind that, if the transition times are too close to each other, the orbits do not appear visually as distinct as shown.
To make the transition times $\tau_{56}$, $\tau_{31}$, $\tau_{64}$ further apart from each other, from
Equations~(\ref{n20choice}), (\ref{parameter_case}), (\ref{tau31}), (\ref{tau56}) and (\ref{tau64}) we see that:
larger $|n_{30}|$ makes $\tau_{64}$ larger, smaller $|n_{10}|$ makes $\tau_{56}$ smaller, and larger $|n_{10} n_{30}|$ makes $\tau_{31}$ larger.
Therefore, one should make $|n_{30}|$ big, $|n_{10}|$ small, and keep $|n_{10} n_{30}|$ of order 1 to keep the three transition times well apart from each other.
This holds for the case $w<1$ below as well.

\begin{figure}[t]
        \begin{center}
        \includegraphics[width=12cm]{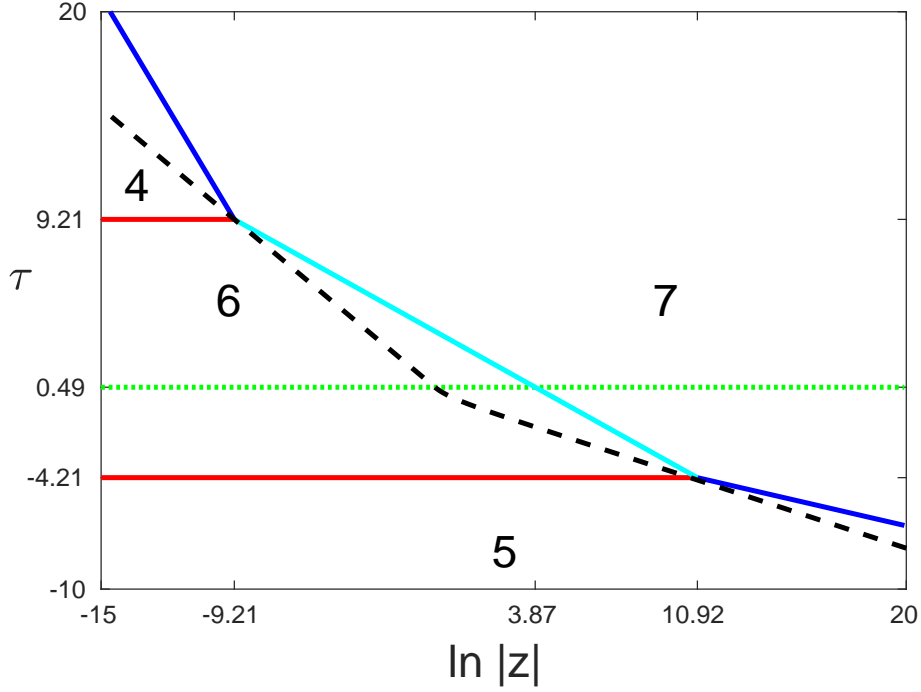}
        \caption{Plot of the cells and transition times for $F$ for Case $w<1$, showing the different scenarios along each fixed $|z|$.
                Each cell is labelled with the index of the dominant term. The parameter values used here are $w=0.5$, $\rho_0=2$, $n_{10} = 0.01$, $n_{30} = 100$.
                $\ln |z|_1 \approx -9.21$, $\ln |z|_2 \approx 3.87$, $\ln |z|_3 \approx 10.92$, $\tau_{56} \approx -4.21$, $\tau_{31} \approx 0.49$. $\tau_{64} \approx 9.21$.
                Dashed line indicates the particle horizon.}
        \label{cell2}
        \end{center}
\end{figure}

\begin{figure}[t]
        \begin{center}
        \includegraphics[30,150][400,250]{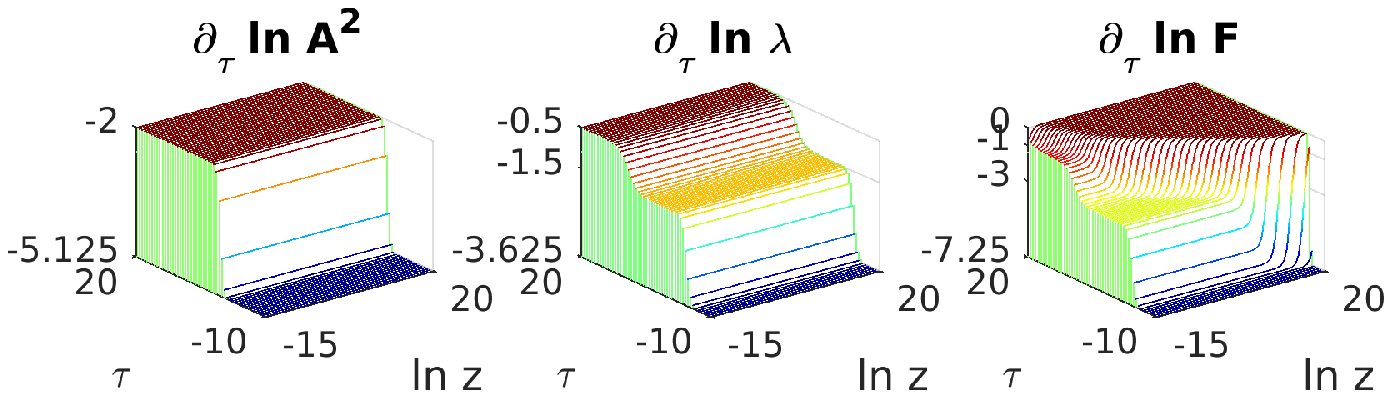}
        \caption{The cascading graphs of $\partial_\tau \ln \mathcal{A}^2$, $\partial_\tau \ln \lambda$ and $\partial_\tau \ln F$, with the same parameter values used in Figure~\ref{cell2}.}
        \label{waterfall_cell2}
        \end{center}
\end{figure}

\begin{figure}[t]
        \begin{center}
        \includegraphics[height=8cm]{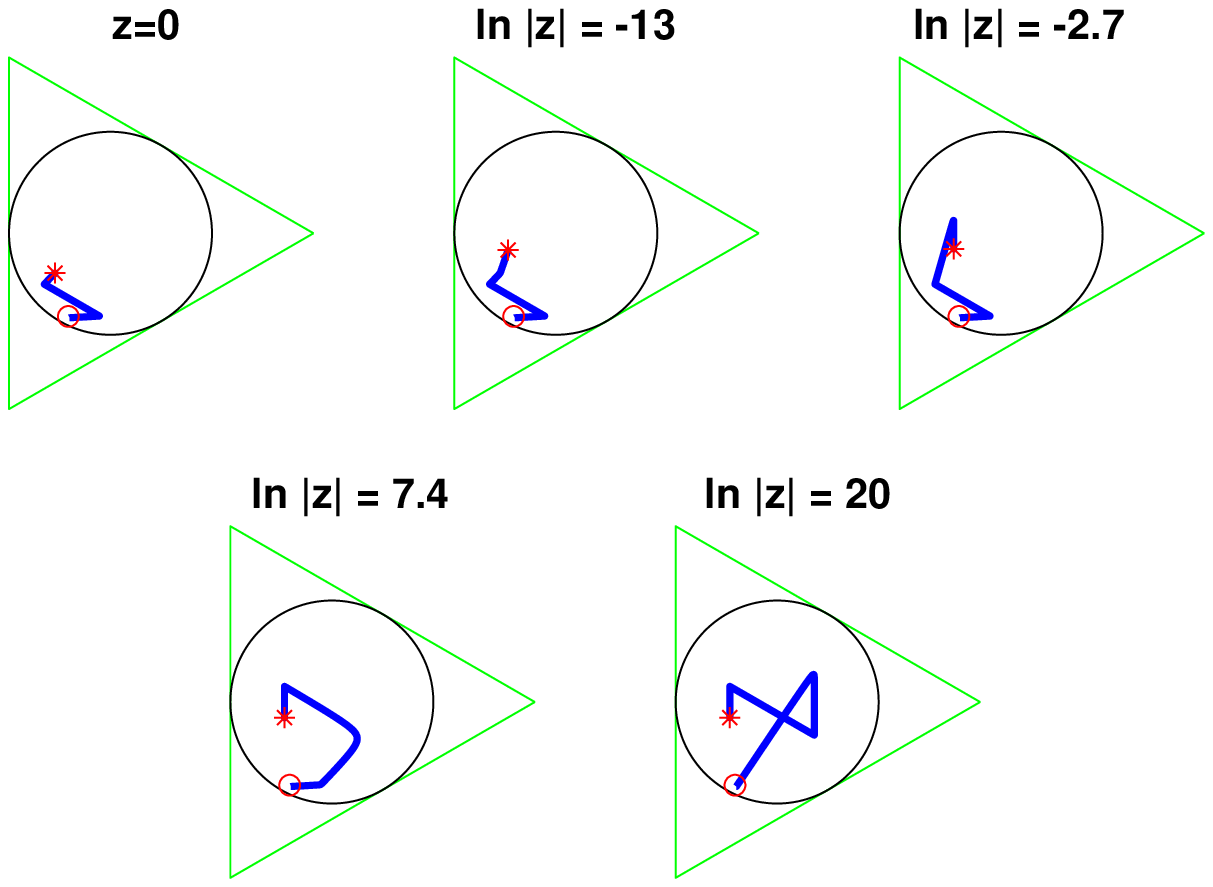}
        \caption{The state space orbits projected onto the Hubble-normalised $(\Sp,\Sm)$ plane, showing a distinctive orbit for each of the five groups of worldlines.
                The parameter values used here are $w=0.5$, $\rho_0=2$, $n_{10} = 0.01$, $n_{30} = 100$. The representative worldlines used are
                $z=0$ and $\ln |z| = -15,\ -2.7,\ 7.4,\ 20$. A red circle marks the start of the orbit, a red star marks the end.
                The orbit along $z=0$ ends at a different point from the rest.}
        \label{orbits2}
        \end{center}
\end{figure}

\subsubsection{Case $w<1$}

For the case $w < 1$, we have $S_5 < S_6 < S_4 < S_7$.
Focussing on the joint spike transition, along worldlines with large enough $|z|$, $T_7$ will become dominant before $T_6$ and $T_4$ do, so the scenario in $F$ is simply $T_5 \to T_7$.
In the other extreme, along $z=0$, the term $T_7$ vanishes, causing a permanent spike to form along $z=0$ at late times (see~\cite{art:ColeyLim2005}).
See Figure~\ref{cell2}. Each cell is represented by a Jacobs equilibrium point.
The parameter values used here are
\be
        w=0.5,\ \rho_0=2,\ n_{10} = 0.01,\ n_{30} = 100,
\ee
giving
\begin{gather}
         \ln |z|_1 \approx -9.21,\ \ln |z|_2 \approx 3.87,\ \ln |z|_3 \approx 10.92,
\notag\\
	 \tau_{56} \approx -4.21,\ \tau_{31} \approx 0.49,\ \tau_{64} \approx 9.21.
\end{gather}
A notable qualitative difference from Figure~\ref{cell1} is that the transition time $\tau_{47}$ has negative slope.
The width of the spike is super-horizon at all times.
The cascading graphs of $\partial_\tau \ln \mathcal{A}^2$, $\partial_\tau \ln \lambda$ and $\partial_\tau \ln F$ are plotted in Figure~\ref{waterfall_cell2}.
Figure~\ref{orbits2} show the distintive orbits for each of the five groups of worldlines.
The orbit along the permanent spike worldline $z=0$ ends at a different Jacobs equilibrium point from the rest.

\section{Discussion}

We have used the technique developed in Moughal's doctoral thesis to analyse the joint spike transition described by the non-OT $G_2$ stiff fluid spike solution,
revealing new groups of worldlines which undergo distinct transitions, and correcting misconceptions about spikes. 
We now know that for transient spike, the spike worldline is not the sole member of its group.
The analysis also reveals the cell-like partition of spacetime of the solution.
This has given us new understanding of the nature of inhomogeneous spacetimes near spacelike singularities.
The discovery also has a practical application.
The detailed transition times and the horizon size estimate may be used to help fine-tune the numerical simulations of $G_2$ spacetimes in fluid-comoving gauge~\cite{art:ColeyLim2016}.

\enlargethispage{1cm}
\section*{Appendix}

The coordinate size of the particle horizon in the $z$ direction is given by the formula
\be
	z_h = \int_\tau^\infty |N|\e^{b^3}\ \d\tau,
\ee
where $|N|\e^{b^3}$ is the coordinate speed of light in the $z$ direction.
For the exact spike solution, $|N|\e^{b^3}$ simplifies to $\mathcal{A}$.
For the joint spike transition with parameter case~(\ref{parameter_case}), the scenario is $T_3 \to T_1$.
So an approximate $z_h$ can be computed:
\be
\label{particle_horizon}
	z_h \approx 	\begin{cases}
				\e^{-\tau}	& \text{for $\tau \geq \tau_{31}$} 
			\\
				\frac{4|n_{10} n_{30} - n_{20}|}{(w+1)^2+4\rho_0}\left[ \e^{-\frac14[(w+1)^2+4\rho_0]\tau} - \e^{-\frac14[(w+1)^2+4\rho_0]\tau_{31}}\right] + \e^{-\tau_{31}}
						& \text{for $\tau < \tau_{31}$.}
			\end{cases}
\ee
This formula is used to plot the particle horizon in Figures~\ref{cell1} and~\ref{cell2}.

\bibliography{}

\end{document}